\documentstyle[epsf,referee]{mn}
\newcommand{\un}[2]{\mbox{\rm\thinspace #1$^{#2}$}}

\newcommand{\be}[1]{\begin{equation}\label{#1}}
\newcommand{\ee}{\end{equation}}

\newcommand{\gsim}{\mathrel{\hbox{\rlap{\lower.55ex \hbox {$\sim$}}
                   \kern-.3em \raise.4ex \hbox{$>$}}}}
\newcommand{\lsim}{\mathrel{\hbox{\rlap{\lower.55ex \hbox {$\sim$}}
                   \kern-.3em \raise.4ex \hbox{$<$}}}}

\newcommand{\sub}[1]{_{\rm #1}}

\newlength{\ffh}

\newcommand{\bea}{\begin{eqnarray}}
\newcommand{\eea}{\end{eqnarray}}
\begin{document}
\title[Shocked by GRB\,970228]
    {Shocked by GRB\,970228:\\
     the afterglow of a cosmological fireball}
\author[Wijers, Rees, and M\'esz\'aros]
       {Ralph A.M.J. Wijers$^1$, Martin J. Rees$^1$, 
        and Peter M\'esz\'aros$^2$\\
        $^1$Institute of Astronomy, Madingley Road, Cambridge CB3 0HA, UK\\
        $^2$Department of Astronomy and Astrophysics, Pennsylvania State 
            University, University Park, PA 16802, USA\\
	E-mail: {\tt ramjw@ast.cam.ac.uk}, {\tt mjr@ast.cam.ac.uk},
	        {\tt nnp@astro.psu.edu} }
\date{\underline{Submitted to MNRAS, 15 April 1997}}
\maketitle

\begin{abstract}
The location accuracy of the BeppoSAX Wide Field Cameras and acute
ground-based followup have led to the detection of a decaying afterglow in
X rays and optical light following the classical gamma-ray burst
GRB\,970228. The afterglow in X rays and optical light fades as a power law
at all wavelengths. This behaviour was predicted for a relativistic blast
wave that radiates its energy when it decelerates by ploughing into
the surrounding medium. Because the afterglow has continued with unchanged
behaviour for more than a month, its total energy must be of order
$10^{51}\un{erg}{}$, placing it firmly at a redshift of order 1.  Further
tests of the model are discussed, some of which can be done with available
data, and implications for future observing strategies are pointed out. We
discuss how the afterglow can provide a probe for the nature of the burst
sources.
\end{abstract}



   \section{Introduction}
   \label{intro}

Gamma-ray bursts are such a mystery in large part because in the more than
three decades since the first report on them (Klebesadel, Strong, \& Olson
1973)\nocite{kso:73} they have remained invisible in any radiation other
than X and $\gamma$ rays. This situation ended in the early hours of
February 28, when the Gamma-Ray Burst
Monitor (40--1000\,keV) and one Wide Field Camera (2--32\,keV) on board the
Italian-Dutch Satellite per Astronomia a Raggi X (BeppoSAX) triggered
on a moderately bright gamma-ray burst and imaged it in X rays to give
an error box only 6 arcmin across (Costa et~al.\ 1997)\nocite{cffzn:97}.
It was still detected at 
a much fainter level by the MECS and LECS X-ray telescopes on BeppoSAX
8 hours after the burst, and again 3 days later, tightening the error box
radius to under an arcmin. Optical images taken with the
William Herschel and Isaac Newton Telescopes at La Palma starting only
20 hours after the burst reveal a fading source that is undoubtedly associated
with GRB970228 (Van Paradijs et~al.\ 1997)\nocite{pggks:97}, and therefore
its location is now known with sub-arcsecond precision. 

Initial reports of an underlying host galaxy of roughly $R=24$ (Van
Paradijs et~al.\ 1997) appear inconsistent with later HST images that show
only a very faint, if any, underlying object (Sahu et~al.
1997a,b)\nocite{slpm:97,slpmp:97}.  The issue of a host and its
implications for the distance scale therefore remain to be resolved, and we
shall be primarily concerned here with the evolution of the gamma-ray burst
itself.  We shall assume that its
contribution to all optical detections thus far is not large.

In Section~\ref{blast} we briefly review the basic properties and
predictions of a blast wave.  In Section~\ref{test} we discuss the light
curves of GRB\,970228 and GRB\,970402 and their afterglow and show that they
agree very well with the predictions of the fireball model (Rees
\& M\'esz\'aros 1992, M\'esz\'aros \& Rees
1997a)\nocite{rm:92,mr:97}. We then briefly discuss the distance scale
in relation to our results and some possible complications due to less
simple blast wave models (Sect.~\ref{discu}). We 
summarise our findings and discuss some implications for observing
strategies in Section~\ref{conclu}.

   \section{The simplest blast wave and remnant evolution}
   \label{blast}

The simplest fireball remnant model is given by the diminishing emission
of a forward blast wave moving ahead of a fireball, which continues to
plough into an increasing amount of external matter beyond the
deceleration radius at which the bulk Lorentz factor $\Gamma$ first
dropped substantially giving rise to the GRB 
(Rees \& M\'esz\'aros 1992, Vietri 1997\nocite{rm:92,vietr:97}). Beyond
this, the Lorentz factor decays as $\Gamma \propto r^{-3/2} \propto
t^{-3/8}$, where $t$ is the observer-frame time (M\'esz\'aros \& Rees
1997a\nocite{mr:97}), and the spectrum is due to synchrotron radiation,
from electrons accelerated to a power law $N(\gamma) \propto \gamma^{-p}$
above the minimum energy $\gamma\sub{m} \propto \Gamma$ imparted by the shock.
The comoving intensity or energy spectrum is
\bea
I_\nu \propto 
\left\{ \begin{array}{ll}
\nu^{\alpha'} & \mbox{for $\nu < \nu\sub{m}$} \\
\nu^{\beta'}  & \mbox{for $\nu > \nu\sub{m}$}
\end{array}
\right.
\label{eq:inu}
\eea   
where the synchrotron break frequency $\nu\sub{m} \propto \Gamma B'
\gamma\sub{m}^2$, and $\beta'=(1-p)/2$.
Observationally (Band et~al.\ 1993)\nocite{bmfsp:93}, $\alpha'=\alpha+1$
is around 0 (single $B'$ synchrotron would give 1/3 but sampling a range
of $B'$ and $\gamma\sub{m}$ can easily give a range around that), while
$\beta'=\beta+1$ is around $-1$ (ranging approximately from $-0.5$ to
$-2$). This is the canonical GRB spectrum in $\gamma$-rays, the observed
$h \nu\sub{m}$ being typically in the range of 50\,keV to 2\,MeV.

The comoving-frame equipartition magnetic field is $B' \propto \Gamma \sim
t^{-3/8}$, so the break frequency drops in time as $\nu\sub{m} \propto \Gamma
B' \gamma\sub{m}^2 \propto \Gamma^4 \propto t^{-3/2}$. At the same time, since
the comoving electron density $n'\sub{e} \propto \Gamma$ and the comoving
width of the emission shell $\Delta R \sim r/\Gamma \sim t^{5/8}$, the
comoving intensity $I'_\nu \propto n'\sub{e} B'^2 \gamma\sub{m}^2 \Delta R/(B'
\gamma\sub{m}^2) \propto n'\sub{e} B' \Delta R \propto t^{-1/8}$, so the
observed flux as a function of observer time is (M\'esz\'aros \& Rees
1997a\nocite{mr:97}) $F_{\nu\sub{m}} \propto t^2 \Gamma^5 I'_{\nu\sub{m}} \propto t^0
\sim$ constant. The observed break frequency $\nu\sub{m}$ crosses the X-ray or
optical band at a time $t\sub{X,op} \sim (\nu_\gamma /\nu\sub{X,op}
)^{2/3} t_\gamma$ after the break frequency lies at $\nu_\gamma$. For
$h\nu\sub{\gamma,X,op}=$ 100\,keV, 5\,keV , and 2.3\,eV, respectively, one
has $t\sub{X}=7t_\gamma$ and $t\sub{op}=170t_\gamma$.

If the time-averaged flux at $\gamma$ rays of the GRB was $F_\gamma$, then
for $t < t\sub{X,op}$ we have $F\sub{X,op} = F_\gamma
(\nu\sub{X,op}/\nu\sub{m})^{\alpha'}$, After $\nu\sub{m}$ crosses the
X-ray or the optical band, we have
\bea 
F\sub{X,op} \propto F_{\nu\sub{m}}(\nu\sub{X,op}/\nu\sub{m} )^{\beta'}\propto 
  t^{(3/2)\beta'}\equiv t^\delta, \mbox{ for $t \geq t\sub{X,op}$ }
\label{eq:fnu}
\eea

The expansion $\Gamma\propto r^{-3/2} \propto t^{-3/8}$ lasts until
the remnant becomes nonrelativistic, after which the remnant enters a
Sedov-Taylor phase. Here $r\propto t^{2/5},~B'\propto t^{-3/5},~\gamma \propto
t^{-6/5},~n'\sub{e} \Delta R \propto t^{2/5}$ and $\nu\sub{m} \propto t^{-3}$,
$F_{\nu\sub{m}} \propto t^{3/5}$, so that the optical (and X-ray, if it were 
detectable) flux has $\nu > \nu\sub{m}$ and goes as $F\sub{op} \propto 
t^{(3+15\beta')/5} \sim t^{-12/5}$ for $\beta'\sim-1$. 
Such a break in behaviour would occur 
when the blast wave has swept up a rest mass energy equal to its initial
energy, at time
\begin{equation}
  \label{eq:tnr}
  t\sub{nr}  =  \left(\frac{3E}{4\pi nm\sub{p}c^5}\right)^{1/3} 
             \!\!\!\simeq  1\un{yr}{}\left(\frac{E_{51}}{n_{-1}}\right)^{1/3}
             \!\!\!\simeq  1\un{d}{}\left(\frac{E_{41}}{n_{-3}}\right)^{1/3},
\end{equation}
where $E$ is the initial explosion energy and $n$ is the density of the
surrounding medium, and
the two numerical expressions are scaled to typical cosmological and
halo cases, respectively.
Such a break has, so far, not been seen in GRB\,970228. Since it has now
been followed for over a month, this firmly places it well beyond the halo
of our Galaxy, if the blast wave model is indeed valid.

   \section{Confrontation with the data}
   \label{test}

      \subsection{Data on the light curve}
      \label{test.data}

In Table~\ref{ta:flux} are listed all the fluxes from gamma-ray to radio
that have been reported since the initial trigger.  All upper limits are
3$\sigma$. Optical and near-infrared magnitudes were translated into fluxes
assuming they were all calibrated on the Vega system. Of course, since many
reports are preliminary, the final calibrated values may differ somewhat
from the table values, but since our emphasis is on trends of the
fluxes when they change by a few orders of magnitude, these corrections
will not affect the results reported here. The foreground reddening at the
location of the burst is $E(B-V)=0.14$ (Burstein \& Heiles
1982)\nocite{bh:82}, so the magnitudes were de-reddened using $A_B=0.8,
A_V=0.4, A_R=0.3, A_I=0.2, A_{J,H,K}=0$.
\begin{table}
\begin{center}
   \caption[]{Fluxes of GRB970228 and its fading counterpart. Time is
              measured from the burst trigger (February 28.124 UT). 
              }
   \label{ta:flux}
\begin{tabular}{@{}cr@{.}lrclll@{}} \hline
$\log t$ & \multicolumn{2}{c}{$F_\nu$}  
                  & $\log\nu$ & band & instrument & ref. \\
 (s)  & \multicolumn{2}{c}{($\mu$Jy)} & (Hz) & &&\\ \hline
 0.30 & 5400&     & 19.38  & $\gamma$ & SAX, TGRS     & 1,2 \\
 0.30 & 6600&     & 18.08  & X        & SAX           & 1,3 \\
 4.17 & $<$0&032  & 19.38  & $\gamma$ & OSSE          & 4 \\
 4.46 &    0&15   & 18.08  & X        & SAX           & 5\\
 4.57 & $<$0&050  & 19.38  & $\gamma$ & OSSE          & 4 \\
 4.86 & $<$700&   &  9.70  & radio    & Westerbork    & 6 \\
 4.88 &   16&8    & 14.74  & V        & WHT La Palma & 11 \\
 4.88 &   17&1    & 14.56  & I        & WHT La Palma & 11 \\
 5.15 & $<$350&   &  9.70  & radio    & Westerbork    & 6 \\
 5.10 & $<$0&013  & 19.38  & $\gamma$ & OSSE          & 4 \\
 5.36 & $<$350&   &  9.70  & radio    & Westerbork    & 6 \\
 5.41 &    2&4    & 14.84  & B        & ARC 3.5m      & 7 \\
 5.49 &    0&0075 & 18.08  & X        & SAX           & 5 \\
 5.61 & $<$1&2    & 14.74  & V        & NOT           & 3 \\
 5.72 & $<$7&5    & 14.56  & I        & Palomar 1.5m  & 8 \\
 5.73 &    1&0    & 14.64  & R        & Keck II       & 8 \\
 5.77 & $<$3600&  & 10.94  & mm       & BIMA          & 9 \\
 5.79 &    0&0043 & 18.08  & X        & ASCA GIS/SIS  & 10 \\
 5.83 & $<$4&0    & 14.64  & R        & INT La Palma  & 13 \\
 5.88 & $<$1&3    & 14.64  & R        & INT La Palma  & 13 \\
 5.88 & $<$0&9    & 14.74  & V        & INT La Palma  & 13 \\
 5.93 &    0&50   & 14.84  & B        & INT La Palma  & 12 \\
 5.93 &    1&0    & 14.64  & R        & INT La Palma  & 12 \\
 6.05 &    1&2    & 14.64  & R        & ESO NTT       & 12 \\
 6.18 &    6&5    & 14.38  & J        & Calar Alto 3.5m & 14 \\
 6.18 &$<$10&7    & 14.26  & H        & Calar Alto 3.5m & 14 \\
 6.18 &$<$10&7    & 14.13  & K        & Calar Alto 3.5m & 14 \\
 6.35 &    0&29   & 14.74  & V        & HST           & 15 \\
 6.35 &    0&62   & 14.56  & I        & HST           & 15 \\
 6.41 &    1&1    & 14.13  & K        & Keck I        & 17 \\
 6.42 &    0&65   & 14.38  & J        & Keck I        & 17 \\
 6.50 &    0&44   & 14.64  & R        & Keck II       & 18 \\
 6.52 &    0&22   & 14.74  & V        & HST           & 16 \\
 6.52 &    0&43   & 14.56  & I        & HST           & 16 \\ \hline
\end{tabular}\\
{\small
\begin{tabular}{@{}p{\columnwidth}@{}}
 (1) Costa et~al. 1997a\nocite{cffzn:97}; 
 (2) Palmer et~al.\ 1997\nocite{pcgkr:97};
 (3) van Paradijs et~al.\ 1997\nocite{pggks:97};
 (4) Matz et~al.\ 1997\nocite{mmgs:97};
 (5) Costa et~al.\ 1997b\nocite{cfpcf:97}; 
 (6) Galama et~al.\ 1997\nocite{gsphb:97};
 (7) Margon et~al.\ 1997\nocite{mdlc:97};
 (8) Metzger et~al.\ 1997b\nocite{mkdgs:97};
 (9) Smith et~al.\ 1997\nocite{slgl:97};
(10) Yoshida et~al.\ 1997\nocite{ykoti:97};
(11) Groot et~al.\ 1997b\nocite{ggpst:97};
(12) Groot et~al.\ 1997c\nocite{ggpms:97};
(13) Tanvir \& Bloom 1997\nocite{tb:97};
(14) Klose et~al.\ 1997\nocite{kst:97};
(15) Sahu et~al.\ 1997\nocite{slpm:97};
(16) Sahu et~al.\ 1997\nocite{slpmp:97};
(17) Soifer et~al.\ 1997\nocite{snamk:97};
(18) Metzger et~al.\ 1997a\nocite{mcbkd:97}.
\end{tabular}
}
\end{center}
\end{table}

The X- and $\gamma$-ray data need a somewhat more careful treatment, since
they are taken over much broader ranges of photon energy. TGRS reported
clear evidence for a break in the spectrum, at $E=100-150\,$keV if an
exponential cutoff was used (Palmer et~al.\ 1997)\nocite{pcgkr:97}. On the
other hand, the GRBM on SAX reported significant flux above 600\,keV, so
the spectral change is more likely to take the form of a power law spectrum
that changes slope at some break energy $E\sub{m}$, like the model of Band
et~al.\ (1993). The four fluxes reported for the initial strong 4-s spike
(which appears contain the bulk of the burst fluence) are (in units of
$10^{-6}\un{erg}{}\un{cm}{-2}\un{s}{-1}$, with subscripts giving the energy
range in keV): $F_{40-200}=1.7$ (Palmer
et~al.\ 1997), $F_{1.5-7.8}=0.1$, $F_{40-600}=4$, $F_{40-1000}=6$ (Van
Paradijs et~al.\ 1997)\nocite{pggks:97}.  Assuming a spectrum that is flat
below $h\nu\sub{m}=E\sub{m}$ ($\alpha'=0$) and has slope $\beta'$ above 
it (eq.~\ref{eq:inu}), we can try to see which $(\beta',E\sub{m})$ 
match the fluxes best.
$E\sub{m}$ must be greater than about 40\,keV and $\beta'$ in the range
$-1$ to $-0.5$ to get a satisfactory fit. Since we know from the TGRS
spectrum that $E\sub{m}\lsim150$\,keV, this pins the parameters down
reasonably well. To specify the X- and $\gamma$-ray fluxes, we use
$E\sub{m}=40$\,keV and $\beta'=-0.8$. The
gamma-ray flux is then given as the fitted value at 100\,keV, 
$F\sub{100\,keV}=4200\,\mu$Jy.  The
OSSE upper limits were likewise translated into values at $100\,$keV.

\begin{figure*}
  \begin{minipage}[b]{0.8\textwidth}
   \epsfxsize=\columnwidth\epsfbox{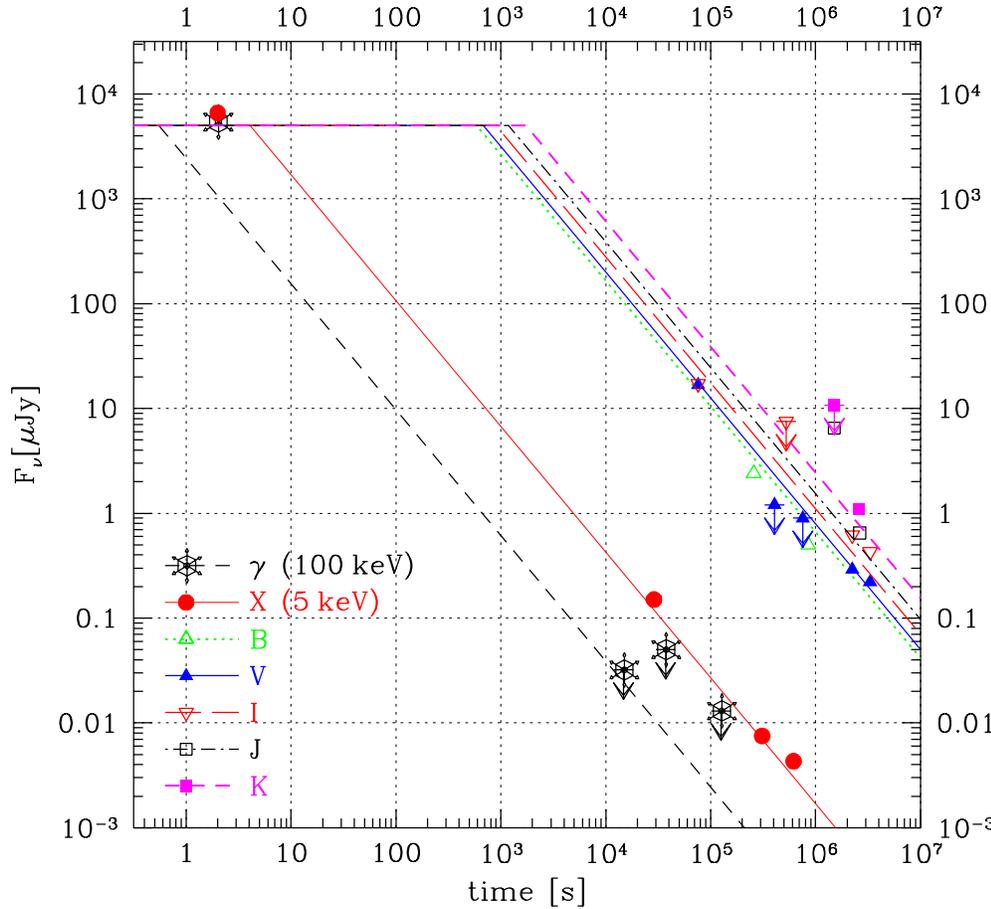}
  \end{minipage}\hfill\begin{minipage}[b]{0.19\textwidth}
\caption{The light curves of GRB\,970228 from $\gamma$ rays to
         near-infrared. The lines indicate the prediction for a relativistic
         blast wave with $\beta'=-0.8$ and $t_V=600$\,s.
         \label{fi:allflu}
         }
  \end{minipage}
\end{figure*}

The initial X-ray flux also follows from the spectral fit, and all other
reported X-ray fluxes are translated into fiducial $F_\nu$ values at 5\,keV
as well. For X rays, this is unlikely to introduce an error of more than
30\%, for gamma rays it may be a bit more. Accounting for uncertainties in
reddening and the preliminary calibration, most optical points should have
errors under 30\%. This could affect the flux offset between
light curves at different bands by up to that amount, but is unlikely to
affect the inferred rate of decay in any single band by much, since the
reddening corrections do not affect them. Moreover, in cases where more
than one measurement in the same band by the same instrument is available,
which presumably suffer from calibration effects in the same manner, the
decay rate from that subset is quite consistent with that of all data in
that band. The flux history of GRB\,970228 
at various photon energies is plotted in Fig.~\ref{fi:allflu}.

      \subsection{Comparing GRB\,970228 and 970402 with the blast wave model}
      \label{test.comp}

Fig.~\ref{fi:allflu} clearly confirms the prediction that the flux of
GRB\,970228 should decline as a power law, in bands where more than 2
measurements well separated in time are available (X-ray, $V$, $I$).
Moreover, a fit with free slope to data in those bands shows that the
exponent of the decay, $\delta$, is the same for all, as the model demands:
$\delta=-1.2$. But then we predict the slope of the spectrum,
$\beta'=2\delta/3=-0.8$. The spectrum in the decaying part of the light
curves (i.e.\ above the break) also follows from the flux ratios at fixed time,
i.e.\ from the vertical offsets between the fitted lines.  This independent
measurement of the spectrum gives $\beta'=-0.78$, remarkably close to the
prediction.

Since after fixing the vertical level of the fit to the $V$ data all
other levels are fixed in the model, we can show the predictions that
follow from it for the other frequencies (Fig.~\ref{fi:allflu}). The agreement
is quite good all the way from X rays to $K$ band. It is too early to say
whether the occasional exception (e.g.\ the first $J$ point) constitutes
interesting extra behaviour. Overall, the agreement in such detail can  be
regarded as a strong confirmation of the model.

It is undecided whether the initial gamma-ray burst derives from the
same mechanism as the afterglow, i.e.\ is the initial part of the blast wave
deceleration, or has a separate origin, e.g.\ internal shocks in the
relativistic wind (Paczy\'nski \& Xu 1994, Rees \& M\'esz\'aros 
1992)\nocite{px:94,rm:94} before the deceleration by the external medium begins.
There is some indication in the $\gamma$-ray light curve of GRB 970228 that
speaks against the former option. The leftmost dashed curve in
Fig.~\ref{fi:allflu} is the prediction for the $\gamma$-ray flux assuming it
originates from the blast wave itself (and can therefore be predicted from
the $V$ and X-ray light curve). It only just falls below the OSSE upper
limits, and even then the break happens too early, as evidenced by the
fact that the initial $\gamma$-ray flux lies above the curve. 
It is therefore possible 
that the gamma rays have a different origin. In that case, we
can no longer rely on the initial $\gamma$-ray flux to set the level of
emission from the blast wave before the break. This implies that the
initial $V$ magnitude of the burst could have been rather low, limited only
by the first measurement ($V=20.9$), rather than 
a (model dependent) value related to the initial $\gamma$-ray flux.

The sparser data on GRB\,970402, which was triggered on April 2.93 UT
(Feroci et~al.\ 1997)\nocite{fcpfz:97,hzcf:97} also have some bearing on
this matter. This burst lasted about 100 seconds with apparently no
obviously dominating initial peak, and its peak flux was about 10 times
lower than GRB\,970228.  It is the second GRB of which afterglow was
detected, again 8 hours after the trigger, but a second attempt after 1.7
days only found an upper limit (Piro et~al.\ 1997)\nocite{pfcaf:97}.
Followup with 1-m telescopes at Siding Spring and SAAO (the burst is
at declination $-69^\circ$) 0.7--1.9 days after trigger place upper limits
of $R=21$ on any fading counterpart (Groot et~al.\ 1997a)\nocite{ggpks:97}.
From the X rays, and initial gamma rays, we once again infer that this
burst has roughly the canonical spectrum of slope 0 below and slope $-1$ to
$-0.8$ above the break, so we expect it to fade in time as a power law with
slope $\delta=$ $-1.5$ to $-1.2$. If we assume all emission is from the
blast wave, and hence the flux at all wavelengths starts at the same value
as the initial $\gamma$-ray flux, then the time at which the break lies at
5\,keV follows from the X-ray detection: $t_5=$ 4--40\,s. The break should
then arrive at $R$ after $t_R=$800--8000\,s, consistent with the
$R$ limit. However, the predicted time for the gamma-ray break becomes
$t_{100}=$0.5--5\,s, rather early for a burst that lasts over 100\,s and is
not dominated by an initial spike, as was GRB\,970228. This discrepancy in
the $\gamma$-ray timing, similar to the case of GRB\,970228, may lend some
support to the independent origin of the initial $\gamma$-ray emission.

To distinguish between these cases, measurements in the first 15 minutes in
X rays or in the first few hours in optical are needed. This could be
achieved with present instruments because positions with 10 arcmin accuracy
become available from the Wide Field Cameras on board BeppoSAX on a 1--3
hour time scale, and a BACODINE-type alert system could spread this
knowledge to observatories around the world in seconds. Also, since
re-pointing the satellite is not done until after some time, the WFC data
immediately following the main burst could be a sensitive diagnostic of
what the initial level of the X-ray afterglow might be.

      \section{Discussion}
      \label{discu}

      \subsection{Galactic models}
      \label{discu.galac}

Since they have much less energy than the cosmological ones,  conventional
matter-dominated fireball blast waves should fade in about a day
(Sect.~\ref{blast}), and thus cannot give rise to the observed long-term
afterglow.

In a galactic-halo model, the X-ray emission could be due to emission from
a surface initially heated to $T\gsim 10^7$ K, which might cool as a power
law in time (regulated by some diffusion time dependent on the depth of
deposition of the energy). The optical emission is much larger than the
$F\sub{op} \sim 10^{-4} F\sub{X}$ expected from the Raleigh-Jeans tail of
such a quasi-thermal surface emission. It is conceivable that the optical
could be a separate component, perhaps associated with reprocessing of
harder radiation by circumstellar material, in which case one might expect
the optical first to increase, peak, and then fade, as the irradiating
``hard" photons first cool to become UV and then become themselves
optical, dropping off exponentially after that.

The Comptonisation model for gamma-ray bursts in the Galactic halo (Liang
et~al.\ 1997)\nocite{lksc:97} predicted the optical brightness 20\,hr
after the burst to be $V=25$ from the observed X-ray flux (Smith
et~al.\ 1997)\nocite{slgl:97}, 4 magnitudes fainter than observed. Hence
it is inconsistent with the behaviour of GRB\,970228. This is due mainly
to the fact that it predicts that the spectrum below the break fades
rapidly as well, in contrast with the blast wave model.

The radiation efficiency in galactic halo blast wave models would
generally be low, $10^{-3}$ or less, due to the smaller magnetic field
strengths at a given time and the shorter expansion time scales at a given
$\Gamma$. Thus one might expect faster turnoff of both the X-ray and
optical emission, due to a decrease of the optical radiation efficiency.
This would be in addition to the steepening caused by moving into the
Sedov-Taylor nonrelativistic phase. The late-time radio afterglow should
come from a large enough region to be resolved by VLBI.

For sources within a few hundred kpc, one could expect the radio emission
to turn on much sooner than in cosmological models, since it is quicker in
reaching a given angular diameter, when the remnant still has sufficient
internal energy to produce observable emission. Furthermore, a galactic
halo neutron star would not be disrupted (since it needs to repeat many
times) and thus could, in principle, produce also coherent radio emission
by pulsar-type mechanisms in the magnetosphere.

      \subsection{Less simple fireball and remnant evolution}
      \label{discu.compl}

The decay model discussed in Sect.~\ref{blast}
is for the simplest external blast wave
scenario, valid for a spherically expanding fireball or a flow which is
uniform inside channels of smaller solid angle. Other types of emission
have been also considered, e.g.\ from the reverse shock moving into the
ejecta, or from ejecta with a frozen-in magnetic field, or from internal
shocks (M\'esz\'aros \& Rees 1997a\nocite{mr:97}). These decay more rapidly
than $F \propto t^{(3/2)\beta'}$ (eq.~\ref{eq:fnu}).  However, there are
additional considerations which can lead to different decay laws,
including some which are slower than $t^{(3/2)\beta'}$.
It is, for instance, probable that the high-$\Gamma$ fireball that gives
rise to the GRB may be beamed (M\'esz\'aros \& Rees 1992, 
1997b)\nocite{mr:92,mr:97b}, and in this case it is
natural for $\Gamma$ to be a function of angle, tapering off towards
the edges of the beam. In a cosmological compact binary disruption scenario,
a central object (presumably a black hole) forms, which is surrounded by
a temporary torus of neutron star debris. A very high bulk Lorentz factor
jet emerges around the rotation axis, which would taper off into a slower
wind, as it mixes with the increasingly baryon loaded wind at large angles
which must result from the super-Eddington radiation from the torus.
As a simple example, we can take a dependence $\Gamma \propto \theta^{-k}$
(M\'esz\'aros \& Rees 1997b)\nocite{mr:97b}, 
where $\theta$ is the angle from the axis of rotation, outside of
the main part of the jet. The deceleration of the fireball commences
at a radius $r\sub{d} \propto \Gamma^{-8/3}$, which is larger for
larger $\theta$. Since radiation can be seen by the observer from within a
cone of angle $\sim \Gamma^{-1}$, the radiation from increasingly larger
angles will be detectable by the observer, at later times which are
given by the deceleration time $t\sub{d}(\theta) \propto \Gamma^{-8/3}$ for
that $\Gamma \sim \theta^{-k}$. This time is of order a day for $\Gamma \sim
10$, and a month for $\Gamma \sim 3$.  In the simplest case where the burst
progenitor supplies equal amounts of energy into equal logarithmic intervals
of $\theta$, the bolometric flux detected as a function of time will then
be $F(t) \propto E(\theta) /t \propto t^{-1}$. If the spectrum is again of
the form $F_\nu \propto [~\nu^{\alpha'},~\nu^{\beta'}~]$ below and above a
time dependent break $\nu\sub{m}$, and we take the field to be a
constant fraction of
the equipartition value, then $\nu\sub{m} \propto \Gamma B' \gamma^2 \propto \Gamma^4
\propto t^{-3/2}$, and $F_{\nu\sub{m}} \sim L /\nu\sub{m} \propto t^{-1} t^{3/2}
\propto t^{1/2}$. If the radiative efficiency does not change in time and
the spectrum evolves homologously, then we have
\bea
F\sub{X,op} \sim
\left\{  \begin{array}{ll}
F_{\nu\sub{m}} (\frac{\nu\sub{X,op}}{\nu\sub{m}})^{\alpha'} 
           & \propto t^{(1+3\alpha')/2} \propto t^{1/2}, \\
 & \hbox{for $t < t\sub{X,op}$,~if $\alpha' \sim 0$}  \\
 F_{\nu\sub{m}} (\frac{\nu\sub{X,op}}{\nu\sub{m}})^{\beta'}  
           & \propto t^{(1+3\beta')/2}  \propto t^{-1}, \\
  & \hbox{for $t > t\sub{X,op}$,~if $\beta' \sim -1$} \\
\end{array}
\right.
\label{eq:fnucomp}
\eea

If the optical radiative efficiency varies in time, the simple homologous
behaviour above introduces additional changes. If we again assume that
fields are a constant fraction of equipartition, then $\nu \propto \Gamma B' \gamma^2
\propto \Gamma^2 \gamma^2$, and the electrons responsible for optical
radiation must satisfy $\gamma\sub{op} \propto \Gamma^{-1}$, so the ratio
of comoving expansion and cooling times is $t'\sub{exp}/t'\sub{cool}
\propto \Gamma^{-2/3} \propto t^{1/4}$. Then the above time power laws
remain the same as long as the optical efficiency $\xi\sub{op} = \max [1,
(t'\sub{exp}/t'\sub{cool})]$ is unity, or they could be flatter by $t^{1/4}$
if the optical efficiency was initially lower than unity and grows (e.g. the
higher energy electrons emitted more efficiently at higher photon energies
than at optical). There are, of course, other possibilities; for instance,
if the field is frozen-in, rather than turbulent, a different $B'$ dependence
needs to be used, etc. The above argument, however, indicates that there
are ways in principle to explain even slower decay time scales than the simple
$\propto t^{(3/2)\beta'}$, as well as faster ones. The afterglow might then 
probe the geometry of the emission, and for suitably oblique viewing angles
one would see afterglow without a burst.

It is important to note that these are testable differences, because the
dependence of the power-law slope of the temporal evolution, $\delta$,
on the slope of the spectrum, $\beta'$, is not the same for eqs.~\ref{eq:fnu}
and \ref{eq:fnucomp}. In the case of GRB\,970228, with $\beta'=-0.8$,
the predictions are $\delta=-1.2$ for the simple case and $\delta=-0.7$ from
eq.~\ref{eq:fnucomp}, so the simple case is favoured for it.

   \section{Conclusion}
   \label{conclu}

As expected, the first detection of a gamma-ray burst in the optical has
greatly furthered our understanding of these enigmatic objects.  We have
found that the simplest fireball model for a gamma-ray burst and its
afterglow agree very well with the data obtained for GRB\,970228 and
GRB\,970402. The longevity of the afterglow argues strongly, within the
context of that model, that GRB\,970228 occurred at substantial
cosmological redshift.  This
raises the interest in unveiling the nature of the faint extended object
coincident with the fading burst. Its faintness would suggest that if it
is a galaxy it probably has $z>1$, whereas the burst itself was moderately
bright. For a no-evolution, standard candle cosmological burst
distribution, its redshift as inferred from the peak flux in $\gamma$ rays
would be smaller, perhaps $z\sim0.3$. This would argue either for a modest
width of the GRB luminosity function, or for significant evolution of
their rate density. Neither is implausible, e.g.\ the star formation rate
evolves so strongly with redshift (Lilly et~al.\ 1996)\nocite{llhc:96}
that if the GRB rate were proportional to it (which a merging neutron star
scenario would naturally demand) the GRB $\log N-\log P$ distribution
would seem Euclidean to $z=1$.

We note that optical flashes perhaps even brighter than
$V \sim 19-20$ from GRB at
cosmological distances (M\'eszaros \& Rees 1997a)\nocite{mr:97}
would be of enormous importance as a tool to study
absorption lines from the early intergalactic medium, features due to the
ISM in the host galaxy, etc.\  (Miralda-Escud\'e, Rees \&
M\'esz\'aros 1997\nocite{mrm:97}), particularly if they arise
from redshifts as large as $z \sim 3-5$. Even at more moderate redshifts,
a good spectrum obtained with a large telescope in the first day could
settle distance scale definitively by showing redshifted absorption
lines or even a Lyman alpha forest\footnote{RAMJW thanks Richard McMahon
for pointing this out.}; even better would be a space-based instrument
with access to the UV part of the spectrum.

We also argue that further significant tests of afterglow models depend
crucially on catching the fading counterpart within the first 0.5 to 3
hours after the trigger. This is feasible in principle with BeppoSAX
since the first (10 arcmin precision) position from the WFC can be derived
within the hour. The history of the first two afterglow detections show that
important information comes from upper limits as well, especially at early
times. So even with modest telescopes rapidly following up a GRB alert is
very worth while. Since the light curve appears to evolve as a power law,
one should try to integrate no longer than the time that passed since the 
GRB trigger in very early measurements, since the brightness of the transient
may change on that time scale.

 \section*{Acknowledgements}

We are grateful to Joshua Bloom, Richard McMahon, Hara Papathanassiou and
Nial Tanvir for 
insightful comments, and for support from a PPARC postdoctoral fellowship
(RAMJW), NASA NAG5-2857, NATO CRG-931446 (PM) and the Royal Society and the
Institute for Advanced Study (MJR).


\end{document}